\documentstyle[12pt]{article}
\begin{document}
\title{Hamiltonian Flow Equations for a Dirac Particle in 
an External Potential
\thanks{Supported by DFG Grant 436 RUS 113/205/1  on
Russian German Cooperation }}
\author{A.B. Bylev$\mbox{}^{\dag}$, H.J. Pirner$\mbox{}^{\dag \dag}$ \\
$\mbox{}^{\dag}$Institute of Physics, St-Petersburg State University \\
$\mbox{}^{\dag \dag}$Institute of Theoretical Physics, Heidelberg University}
\date{\today}
\maketitle
\thispagestyle{empty}

\begin{abstract}
We derive and solve the Hamiltonian flow equations for a Dirac particle
in an external static potential. The method shows a general procedure 
for the set up of continuous unitary transformations to reduce the
Hamiltonian to a quasidiagonal form.
\end{abstract}
\pagebreak
The well known Foldy-Wouthuysen (FW) transformation 
\cite{FW}
is a unitary transformation that separates big and small components in the
Dirac 
equation for a particle interacting with an external
electromagnetic field. If the external field is stationary then the FW 
transformation 
reduces to a unitary
transformation of the Hamiltonian $H$ of the Dirac particle
\begin{equation}
H'=e^{i S} ( \vec{\alpha} (\vec{p} - e \vec{A}(\vec{x})) + \beta m + e A_0
(\vec{x}) ) e^{- i S} 
\label{1}
\end{equation}
where $S$ can be expanded in a sequence of hermitean operators in powers
of $1/m$.
Recently a novel method to  diagonalize Hamiltonians 
via continuous unitary 
transformations has been
proposed \cite{Wegner}, where the Hamiltonian $ H $ and the  $ S $ 
are considered to be 
functions of a continuous variable, the flow parameter $ l $. 
This new method has been applied to several topics like the $n-$ orbital
model \cite {Wegner}, superconductivity and the electron phonon interaction
\cite {Le,Mi}.
We want to apply it here to the Dirac particle in an external field
in order to explore its possibilities with a  well known
problem. Thereby we can gain experience in solving 
problems like the elimination of negative energy states in near light 
cone QCD \cite {Pirner}, where the solution has yet to
be discovered. For infinitesimal steps $\exp (\eta (l) dl)$
a continuous unitary transformation 
of the Hamiltonian $H$ leads to the so-called Hamiltonian flow equation
\begin{equation}
\frac{d H(l)}{d l} = [ \eta (l), H(l)] \label{2}
\end{equation}
with initial condition $ H(0) = H$ . The 
transformation matrix $ \eta (l)$ is antihermitian and related to
$S$ as follows
$$
\eta (l) = (\frac{d}{dl} e^{i S(l)}) e^{- i S(l)}.
$$
The parameter $l$ has dimension of $energy^{-2}$ and
characterizes the amount of diagonalization which has been achieved.
In the energy representation the matrix elements of the
Hamiltonian $H$ will be grouped in a band of width $\Delta E=\frac{1}{\sqrt(l)}$
around the diagonal, if $\eta (l)$
is chosen in the correct way. 
The result of the transformation depends critically on the choice of 
$\eta (l)$ as a function of 
$H(l)$. 
In ref. \cite{Wegner}  the following expression has been proposed:
\begin{equation}
\eta (l) = [H_{D}(l), H(l)] \label{3}
\end{equation}
where $H_{D}(l)$ is the diagonal part of $H(l)$. Indeed such a choice of $\eta 
(l)$ 
leads to a
quasidiagonalization of $H(l)$ when $l \rightarrow \infty$. From 
eqs. 
(\ref{2}) and
(\ref{3}) we obtain that
$$
\frac{d}{dl} \mbox{tr} H^2_{D} (l) = - \frac{d}{dl} \mbox{tr} H^2_{ND}(l) = 2 \mbox{tr} ([H(l),H_{D}(l)]
[H_{D}(l),H(l)]) \geq 0
$$
Thus the sum of the absolute squares of the nondiagonal elements decreases,
whereas the sum of the diagonal matrix elements squared increases with $l$.
However, this line of reasoning only holds
if one can solve eqs. (\ref{2}),(\ref{3}) exactly. For approximate 
solutions of eqs.
(\ref{2}) and (\ref{3}) divergencies can arise\cite{Wegner}.

In this report we want to present a continuous transformation 
of the Dirac Hamiltonian via flow equations
(\ref{2}) the derivation of which is guided by the FW transformation.
The basic idea of the FW transformation is a 
decomposition of the
Hamiltonian
into two noninteracting parts corresponding to positive and negative energies. 
Therefore, the transformed
Hamiltonian must commute with the $\beta$ matrix and we choose as ansatz 
$\eta (l)$ 
in the form
\begin{equation}
\eta (l) = [\beta m, H(l)] \label{4}
\end{equation}

The Hamiltonian $H(l)$ can be presented as a sum of an
even operator ${\cal E} (l)$  and  odd operator ${\cal O}(l)$,
where the even or oddness is defined by the commutation relations
of the respective operators with the $\beta-$ matrix.
\begin{eqnarray*}
H(l) &=& {\cal O}(l) + {\cal E}(l) \label{5},\\
{\cal E}(l) \beta &=&+ \beta {\cal E}(l),\\
{\cal O}(l) \beta &=& - \beta {\cal O}(l).
\end{eqnarray*}
The initial conditions for the evolution equation
$$
\frac{d H(l)}{d l} = [ \eta (l), H(l)] 
$$
are obtained by using the plus commutator $[\vec{\alpha},\beta]_{+}=0$.
\begin{eqnarray*}
{\cal E}(0)&=&\beta m + e A_0,\\
{\cal O}(0)&=&\vec{\alpha} (\vec{p} - e \vec{A}).
\end{eqnarray*}
Then our ansatz for the transformation matrix $\eta(l)$ becomes
$$
\eta (l) = 2 m \beta {\cal O}(l)
$$
The flow equation can be split up into the following system
of two equations
\begin{eqnarray}
\frac{d {\cal E}(l)}{dl} & = & 4 m \beta {\cal O}^2 (l) , \label{7}\\
\frac{d {\cal O}(l)}{dl} & = & 2 m \beta [{\cal O}(l), {\cal E}(l)] . \label{8}
\end{eqnarray}
As we will show now, the system of eqs. (\ref{7}),(\ref{8}) 
can be solved perturbatively 
in 
$1/m$.
It is convenient to introduce a dimensionless flow 
parameter 
$ \lambda=l*m^2$. Since 
${\cal E}(0) = \beta m + e A_0$ the expansion of ${\cal E}
(\lambda)/m$ in a series in $1/m$
contains terms starting with the zeroth  order term
\begin{equation}
\frac{1}{m} {\cal E}(\lambda) = {\cal E}_0 (\lambda) + \frac{1}{m} {\cal E}_1 
(\lambda) + 
\frac{1}{m^2} {\cal E}_2 (\lambda)
+ \ldots , \label{9}
\end{equation}
whereas the expansion of ${\cal O}(\lambda)/m$ starts with the first order
\begin{equation}
\frac{1}{m} {\cal O}(\lambda) = \frac{1}{m} {\cal O}_1 (\lambda) + 
\frac{1}{m^2} {\cal O}_2 (\lambda) 
+ \ldots
\label{10}
\end{equation}
Due to eqs. (\ref{9}) and (\ref{10})  we have
in $n$-th order :
\begin{eqnarray}
\frac{d}{d \lambda} {\cal E}_n (\lambda) & = & 4 \beta \sum_{k=1}^{n-1} 
{\cal O}_k (\lambda) 
{\cal O}_{n-k}(\lambda)
\label{11} \\
\frac{d}{d \lambda} {\cal O}_n (\lambda) & = 
& - 4 {\cal O}_n (\lambda) + 2 \beta \sum_{k=1}^{n-
1} 
[{\cal O}_k (\lambda),
{\cal E}_{n-k} (\lambda) ] \label{12}
\end{eqnarray}
The solution of these equations  are
\begin{eqnarray}
{\cal E}_n (\lambda) &=& {\cal E}_n (0) + 4 \beta \int_0^\lambda d \lambda
^{\prime} (\sum_{k=1}^{n-1} 
{\cal O}_k (\lambda^{\prime}) {\cal
O}_{n-k} (\lambda^{\prime}) )   \label{14} \\
{\cal O}_{n} (\lambda) &=& {\cal O}_{n}(0) e^{- 4 \lambda} +
 2 \beta e^{-4 \lambda} \int_0^\lambda d \lambda^{\prime}
(\sum_{k=1}^{n-1} [e^{4 \lambda^{\prime}} {\cal O}_k (\lambda^{\prime}), 
{\cal E}_{n-k}(\lambda^{\prime})])  
\label{13}
\end{eqnarray}
with initial conditions
\begin{eqnarray*}
&&{\cal E}_{0}(0)=\beta,\ {\cal E}_{1}(0) = e A_0(\vec{x}),\ {\cal E}_{n}(0)=0 
\mbox{ if }n \geq 2 ,\\ 
&&{\cal O}_{1}(0)= \vec{\alpha}(\vec{p} - \vec{A}(\vec{x})),\ 
{\cal O}_{n}(0)=0 \mbox{ if }n \geq 2.
\end{eqnarray*}
One sees that ${\cal O}_{n}(\lambda)$ exponentially goes to zero
when $\lambda \rightarrow \infty$. By 
virtue of this behavior integrals in eq.(\ref{14}) do not diverge and 
${\cal E}_n (\lambda)$ is finite when $ \lambda \rightarrow \infty$.

For the first four orders we have
\begin{eqnarray*}
{\cal E}_{0}(\infty)& = & {\cal E}_{0}(0) = \beta ,\\
{\cal E}_{1}(\infty)& = & {\cal E}_{1}(0) = e A_{0}(\vec{x}),\\
{\cal E}_{2}(\infty)& = & \frac{1}{2} \beta {\cal O}_{1}^2(0) = \frac{1}{2}\beta 
[(\vec{p} - e \vec{A}(\vec{x}))^2 - e \vec{\sigma} \vec{B}(\vec{x})] ,\\
{\cal E}_{3}(\infty)& = & \frac{1}{8} [[{\cal O}_{1}(0), {\cal E}_{1}(0)],{\cal O}_{1}(0)]  = \\
&=& - \frac{i e}{8} \vec{\sigma} (\nabla \times \vec{E}(\vec{x})) - \frac{e}{4} \vec{\sigma} (
\vec{E}(\vec{x}) \times (\vec{p} - e \vec{A}(\vec{x}))) - \frac{e}{8} (\nabla \vec{E}(\vec{x})).
\end{eqnarray*}
These terms reproduce well known Foldy-Wouthuysen Hamiltonian \cite{FW}. The formulae (\ref{14}),(\ref{13}) let us simply get expressions in any order in $1/m$. 

\vspace{1cm}

Similarily to the renormalization group equations the Hamiltonian flow equations
approach the correct solution in infinitesimal steps. We have derived
in this new method a solution of the flow equations
(\ref{2}) and (\ref{3}) 
for the problem of a Dirac particle in an external potential.
This solution has a more general validity. We 
have shown how to choose the diagonalizing matrix $\eta$,
if the Hamiltonian 
can be
represented as a sum of even and odd operators 
under commutation with some operator 
$D$.
Choosing the operator $\eta (l)$ in the form 
$[D, H(l)]$ allows us to transform
the Hamiltonian $H$ to a form which commutes with $D$ whereas the 
anticommuting part 
goes to zero
when the  flow parameter goes to infinity. 

\vskip .5 in
{\bf Acknowlegments} \\
We thank Andreas Mielke for useful discussions. A.B thanks the Deutsche 
Forschungsgemeinschaft for
the support of his visit to the Institute of Theoretical Physics of Heidelberg 
University.

\end{document}